%Paper: hep-ph/9505315
%From: konig@osiris.phy.uqam.ca (Heinz Konig)
%Date: Mon, 15 May 95 23:02:31 -0400
%Date (revised): Mon, 15 May 95 23:33:26 -0400

\magnification=\magstep1
\overfullrule=0pt
\hsize=15.4truecm
{\nopagenumbers \line{\hfil UQAM-PHE-95/01}
\vskip1cm
\centerline{\bf REANALYSIS OF THE MASS DIFFERENCE OF
$B_d^0-\overline B_d^0$\ WITHIN}
\centerline{\bf THE MINIMAL SUPERSYMMETRIC STANDARD MODEL}
\vskip2cm
\centerline{{G. COUTURE AND H. K\"ONIG}
\footnote*{email:couture, konig@osiris.phy.uqam.ca}}
\centerline{D\'epartement de Physique}
\centerline{Universit\'e du Qu\'ebec \`a Montr\'eal}
\centerline{C.P. 8888, Succ. Centre Ville, Montr\'eal}
\centerline{Qu\'ebec, Canada H3C 3P8}
\vskip2cm
\centerline{\bf ABSTRACT}\vskip.2cm\indent
We present a detailed and complete calculation of the
loop corrections to the mass difference $\Delta m_{B_d^0}/m_{B_d^0}$.
We include charginos and scalar up quarks as well as gluinos
and scalar down quarks on the relevant
loop diagrams. We include the mixings of the charginos
and of the scalar partners of the left and right
handed quarks. We find that the gluino contribution to this quantity is
important with respect to the chargino contribution only in a small
part of phase space: mainly when the gluino mass is small ($\sim$ 100 GeV) and
the symmetry-breaking parameter $m_S$ is below 300 GeV. This contribution is
also important for very large values of $\tan\beta$ ($\sim$ 50) irrespective
of the other parameters. Otherwise, the chargino contribution dominates vastly
and
can be roughly as large as that of the Standard Model.
We also present the
contribution of the charged Higgs to the mass difference
$\Delta m_{B_d^0}/m_{B_d^0}$\
in the case $m_b\tan\beta\ll m_t\cot\beta$. This last contribution can be
larger than the Standard Model contribution for small values of the Higgs
mass and small values of $\tan\beta$.
\vskip1cm
\centerline{ May 1995}
\vfill\break}
\pageno=1
{\bf I. INTRODUCTION}\hfill\break\vskip.2cm\noindent
The mass difference $\Delta m_{B_d^0}/m_{B_d^0}\approx
6.4\times 10^{-14}$\ GeV [1] is an experimental well known
value. It was calculated within the standard model,
where W bosons and up quarks run in the
relevant box diagrams, a long time ago [2--5 and references
therein]. These authors pointed out the importance of
the top quark mass in the $B_d^0$\ system, which is due
to the parameters $K_{ij}$\ of the Kobayshi--Maskawa matrix
(in the $K^0$\ system the top quark contribution is smaller
than the charm quark contribution since
$(K_{21}^\ast K_{22})^2m^2_c\ge (K^\ast_{31}K_{32})^2m^2_t$\ whereas
in the $B_d^0$\ system the term $(K^\ast_{31}K_{33})^2m^2_t$\
is dominant). The measured value of
$x_{B_d^0}=\Delta m_{B_d^0}\times \tau_{B_d^0}\approx 0.71$\
was unexpectedly high. As explanation
it was shown that the top quark mass had to be larger than
45 GeV, which was quite surprising at this time. Nowadays
from CDF we know that the top quark mass  is
almost 4 times that much; with a given value of 174 GeV [6].
\hfill\break\indent
Because of such a large top quark mass we have to reconsider
the influence of one of the most favoured models beyond the
SM, its minimal supersymmetric extension (MSSM) [7], to
$\Delta m_{B_d^0}$. A rough estimate of the influence of
the MSSM to the mass difference of the $K^0$\ system was
first done by several authors [8--10 and references therein].
These authors replaced the W bosons and up quarks by their super
partners, the charginos and scalar up quarks. Rough estimate
it was, because they neglected the mixing of the scalar
partners of the left and right handed up quarks or the
mixing of the charginos. Neglecting the mixing of the
charginos in general lead to wrong statements as was shown
in [11] and the importance of the mixing in the scalar top
quark sector due to the heavy top quark can be seen in [12--13].
\hfill\break\indent
In this paper we present a detailed and complete calculation
of the charginos and scalar up quarks contribution to the
mass difference of the $B_d^0$\ system. In the calculation
we neglect all masses of quarks besides the top quark mass.
As explained above our results cannot be used in the $K^0$\
system.
\hfill\break\indent
More than ten years ago it was shown that loop diagrams
induce flavour changing couplings of the gluinos to the
down quarks and their scalar partners [14--15]. Since this
coupling is strong its influence was first analysed in the
$K^0$\ system [15--17] and later in the $B_d^0$\ system
[18--19]. There it was shown that for a small top quark
mass of 40 GeV and masses of the gluinos and scalar
down quarks at the lower experimental limit of
around 60 GeV at this time,
 the SUSY corrections are of the same order
and even higher than the SM result. In this paper we repeat
those calculations
and give more general results: we include the
mixing of the scalar partners of the left and right
handed bottom quarks. This might become important
if $\tan\beta=v_2/v_1\gg 1$\ where $v_{1,2}$\ are the
vacuum expectation values (vev's) of the Higgs particles
in the MSSM.
\hfill\break\indent
In the next section we present the calculation and discuss the
results in the third section. We end with the conclusions.
\vfill\break\noindent
%\hfill\break\vskip.2cm\noindent
 {\bf II. SUSY CORRECTIONS TO THE $B_d^0$\ SYSTEM}\vskip.2cm
In the SM the mass difference in the $B_d^0$\ system
is obtained by calculating the box diagrams, where W bosons
and up quarks are taken within the loop. After summation
over all quarks it turns out that the top quark gives the
main contribution to $\Delta m_{B_d^0}$. The result is
well known and given by [2]:
$$\eqalignno{{{\Delta m_{B_d^0}}\over{m_{B_d^0}}}=&
{{G_F^2}\over{6\pi^2}}f^2_BB_B\eta_tm^2_W
(K_{31}^\ast K_{33})^2S(x_t)&(1)\cr
S(x_t)=&x_t\lbrace {1\over 4}+{9\over 4}(1-x_t)^{-1}
-{3\over2}(1-x_t)^{-2}\rbrace -{3\over 2}{{{x_t}^3}\over
{(1-x_t)^3}}\ln x_t\cr}$$
$f_B$, $B_B$\ are the structure constant and the
Bag factor obtained by QCD sum rules and $\eta_t$\
a QCD correction factor [2].
For a large top quark mass their values are given by
$f_B=0.180$\ GeV, $B_B=1.17$\ and $\eta_t=0.55$ [3].
\hfill\break\indent
To obtain $\Delta m_{B_d^0}$\ within the MSSM we have to
calculate all those diagrams as shown in Fig.1, where we
also include the so called "mass insertion" diagram
denoted with $\otimes$\ [8]. This notation means that in
this diagram $P_R(\rlap/ k+m)P_R=mP_R$\ remains whereas
the other one gives $P_R(\rlap/ k+m)P_L=\rlap/ kP_L$\
($P_{L,R}$\ are the projection operators).
In our calculation we take the full set of couplings as
given in Fig.22 and Fig.23 of [20]\footnote{$^1$}{In Fig.22 b)+d)
and Fig.23 b)+d) $\gamma_5$\ has to be replaced
by $-\gamma_5$}. Furthermore as mentioned above
we include the mixing of the charginos\footnote{$^2$}{
A more detailed description of the mixing of the charginos
and of the scalar top quarks can be found in [21].}
as well as the mixing
of the scalar partners of the left and
right handed quarks, that is instead of the current eigenstates
$\tilde q_{L,R}$\ we
work with the mass eigenstates
$$\tilde q_1=cos\Theta\tilde q_L+\sin\Theta\tilde q_R\qquad
\tilde q_2=-\sin\Theta\tilde q_L+\cos\Theta\tilde q_R\eqno(2)$$
In the scalar top quark sector we have the following matrix [22]:
$$M^2_{\tilde t}=\left(\matrix{m_{\tilde t_L}^2+
m_t^2+0.35D_Z&-m_t(A_t+\mu\cot\beta)\cr
-m_t(A_t+\mu\cot\beta)&m_{\tilde t_R}^2+m_t^2+0.15D_Z\cr
}\right)\eqno(3)$$
Here $D_Z=m_Z^2\cos 2\beta$,
$0.35=T_3^t-e_t\sin\Theta_W$\ and $0.15=e_t\sin\Theta_W$.
$m_{\tilde t_{L,R}}$\ are
soft SUSY breaking mass terms, $A_t$\ the parameter from the
trilinear scalar interaction and $\mu$\ the mixing mass term
of the Higgs bosons.\hfill\break\indent
The mass matrix of the scalar top quark is of importance,
when we consider the scalar up quarks and charginos within
the loop. The mass matrix of the scalar bottom quark would
be of importance when we would take neutralinos and scalar
down quarks within the loop. Here we have to be more careful
since a while ago it was shown in [14--15], that when
including loop effects flavour changing couplings
of the scalar partner of the left handed down quarks with
the gluinos are created, whereas the couplings of the
gluinos with the scalar partner of the right handed down
quarks remain flavour diagonal, that is only the scalar
partners of the left handed
down quarks have to be considered in the
relevant loop diagrams to $B_d^0-\overline B_d^0$\
mixing. For the mass matrix of the scalar bottom quark
we therefore have to take at 1 loop level:
$$M^2_{\tilde b}=\left(\matrix{m_{\tilde b_L}^2+
m_b^2-0.42D_Z-\vert c\vert m^2_t&-m_b(A_b+\mu\tan\beta)\cr
-m_b(A_b+\mu\tan\beta)&m_{\tilde b_R}^2+m_b^2-0.08D_Z\cr
}\right)\eqno(4)$$
with $T_3^b-e_b\sin\Theta_W=-0.42$\ and $e_b\sin\Theta_W=-0.08$.
The model dependent parameter $c$\ plays
 a crucial role in the calculation
of the gluino and scalar down quark contribution to the
mass difference in the $B_d^0$\ system. The value of $c$\
is negative and of order 1 ($\vert c\vert$ increases
with the soft SUSY breaking mass term $m_S$\ and decreases
with the top quark mass [18]). In the following we take
$c=-1$, although we keep in mind that it is more likely smaller in
magnitude.
The mixing term might only get important in the case
$\tan\beta\gg 1$.
\hfill\break\indent
When calculating the box diagrams in Fig.1 we used
the rules given in appendix D in [7].
After a lengthy but straightforward calculation we obtain
the following result when charginos and scalar up quarks
are running on the loop:
$$\eqalignno{{{\Delta m_{B_d^0}}\over{m_{B_d^0}}}=&
{{G_F^2}\over{4\pi^2}}f^2_BB_Bm^4_W(K^\ast_{31}K_{33})^2
\lbrack Z^{\tilde W}_{11}-2 Z'^{\tilde W}_{31}+
\tilde Z^{\tilde W}_{33}\rbrack&(5)\cr
Z^{\tilde W}_{11}=&\sum\limits_{i,j=1,2}V^2_{i1}V^2_{j1}G^{ij}_
{\tilde u\tilde u}\cr
Z'^{\tilde W}_{31}=&\sum\limits_{i,j=1,2}V_{i1}V_{j1}\Bigl\lbrace
V_{i1}V_{j1}\lbrack c^2_{\Theta_t}G^{ij}_{\tilde t_1
\tilde u}+s^2_{\Theta_t}G^{ij}_{\tilde t_2\tilde u}\rbrack
\cr&+{{m^2_t}\over{2m_W^2\sin^2\beta}}V_{i2}V_{j2}\lbrack
s^2_{\Theta_t}G^{ij}_{\tilde t_1\tilde u}+c^2_{\Theta_t}
G^{ij}_{\tilde t_2\tilde u}\rbrack\Bigr\rbrace\cr
\tilde Z^{\tilde W}_{33}=&\sum\limits_{i,j=1,2}\Bigl\lbrace
V^2_{i1}V^2_{j1}\lbrack c^4_{\Theta_t}G^{ij}_{\tilde t_1
\tilde t_1}+2c^2_{\Theta_t}s^2_{\Theta_t}G^{ij}_{\tilde t_1
\tilde t_2}+s^4_{\Theta_t}G^{ij}_{\tilde t_2\tilde t_2}\rbrack\cr
&+{{m^2_t}\over{m_W^2\sin^2\beta}}V_{i1}V_{i2}V_{j1}V_{j2}
\lbrack c^2_{\Theta_t}s^2_{\Theta_t}(G^{ij}_{\tilde t_1
\tilde t_1}+G^{ij}_{\tilde t_2\tilde t_2})+(c^4_{\Theta_t}
+s^4_{\Theta_t})G^{ij}_{\tilde t_1\tilde t_2}\rbrack\cr
&+{{m^4_t}\over{4m_W^4\sin^4\beta}}V_{i2}^2V^2_{j2}
\lbrack s^4_{\Theta_t}G^{ij}_{\tilde t_1\tilde t_1}
+2c^2_{\Theta_t}s^2_{\Theta_t}G^{ij}_{\tilde t_1\tilde t_2}
+c^4_{\Theta_t}G^{ij}_{\tilde t_2\tilde t_2}\rbrack\Bigr\rbrace\cr
G^{ij}_{ab}:=&\tilde F^{ij}_{ab}+2 _M\tilde F^{ij}_{ab}\cr}$$
$c^2_{\Theta_t}=\cos^2\Theta_t$, $s^2_{\Theta_t}=\sin^2\Theta_t$\
and $\sin\beta$\ can be extracted from $\tan\beta$.
$\tilde F^{ij}_{ab}$\
and $_M\tilde F^{ij}_{ab}$\
are given in the appendix A.
$m_{i,j}=m_{\tilde W_{i,j}}$\ are the mass eigenvalues
of the charginos
and $m_{\tilde u, \tilde t_{1,2}}$\
the masses of the scalar up quark and the eigenstates
of the scalar top quark including the mixing.
$V_{ij}$\ and $U_{ij}$\ are the diagonalizing
matrices of the charginos as given in eq. C19 in [7]
taken to be real.
$\tilde F^{ii}_{ab}$\ and $_M\tilde F^{ii}_{ab}$\
are the same functions as given in eq.C.2 in [23].
\hfill\break\indent
Similarly, when gluinos and
scalar down quarks run on the loop we obtain
\footnote{$^3$}{Eq.6 agrees with [18--19]
for $c_{\Theta_b}=1$.}
$$\eqalignno{{{\Delta m_{B_d^0}}\over{m_{B_d^0}}}=&
-{{\alpha_s^2}\over{54}}f_B^2B_B(K^\ast_{31}K_{33})^2
\lbrack Z^{\tilde g}_{11}-2 Z'^{\tilde g}_{31}+
\tilde Z^{\tilde g}_{33}\rbrack&(6)\cr
Z^{\tilde g}_{11}=&T^{\tilde g}_{\tilde d\tilde d}\cr
Z'^{\tilde g}_{31}=&c^2_{\Theta_b}T^{\tilde g}_{\tilde b_1
\tilde d}+s^2_{\Theta_b}T^{\tilde g}_{\tilde b_2\tilde d}\cr
\tilde Z^{\tilde g}_{33}=&c^4_{\Theta_b}T^{\tilde g}_
{\tilde b_1\tilde b_1}+2c^2_{\Theta_b}s^2_{\Theta_b}
T^{\tilde g}_{\tilde b_1\tilde b_2}+s^4_{\Theta_b}
T^{\tilde g}_{\tilde b_2\tilde b_2}\cr
T^{\tilde g}_{ab}:=&11\tilde F^{\tilde g}_{ab}+
4 _M\tilde F^{\tilde g}_{ab}\cr}$$
$\tilde F^{\tilde g}_{ab}$\ and $_M\tilde F^{\tilde g}_{ab}$\
can be obtained by setting $m_i=m_j\leftrightarrow m_{\tilde g}$\
in the functions given in appendix A.
Since we neglected all quark masses
\footnote{$^4$}{In calculating the box diagram it is
safe to neglect the bottom quark mass, since in the
coupling there is no $\tan\beta$\ dependance and
therefore we can use $m_b\ll m_t$.}
 beside the top
quark mass we made use of $Z_{11}=Z_{12}=Z_{21}=Z_{22}$\
and $Z_{13}=Z_{31}=Z_{32}=Z_{23}$.
Since the mixing of the scalar quark is proportional
to the quark masses $c^2_{\Theta_b}=\cos^2\Theta_b\approx
1$, only for large values of $\tan\beta$\ the mixing angle
of the scalar bottom quark mass becomes more important.
\hfill\break\indent
Finally we also want to comment on
the charged Higgs boson
contribution to the mass
difference in the $B_d^0$\ system. In the case of
neglecting bottom quark mass that is
$m_b\tan\beta\ll m_t\cot\beta$\ we obtain [25]:
$$\eqalignno{\Delta m_{B_d^0}/
m_{B_d^0}=&{{G^2_F}\over{16\pi^2}}m^4_t
\cot^4\beta f^2_BB_B(K_{31}^\ast K_{33})^2
\bigl\lbrace \tilde F^{tt}_{H^+H^+}&(7)\cr
&+2\tan^2\beta\lbrack\tilde F^{tt}_{H^+W^+}+4(m_W/m_t)^2
_M\tilde F^{tt}_{H^+W^+}\rbrack\bigr\rbrace\cr}$$
When one has $m_b\tan\beta\sim m_t\cot\beta$\
one should not neglect the bottom quark
mass when calculating the box diagram; this complicates
greatly the calculations.

\hfill\break\indent
\hfill\break\vskip.2cm\noindent
{\bf III. DISCUSSIONS}\vskip.2cm
We now present those contributions  for different values of Higgs, gaugino,
gluino and scalar quark masses. We also vary $tan\beta$ and the
symmetry-breaking scales. As input parameter we take
$m_{\rm top}=174$\ GeV, $m_{\rm b} = 4.5$\ GeV, $\alpha = 1/137$\ and
for the strong coupling
constant $\alpha_s=0.1134$.
For a top quark mass of 174 GeV the SM result eq.1 gives
a value of $4.67\times 10^{-16}$.
\hfill\break\indent
We first show on fig.2 the charged Higgs contribution. We see that for small
values of $\tan\beta$ and light Higgs, this contribution can exceed that of
the SM. For $\tan\beta = 1$, even for very large Higgs masses, this
contribution is still 20\% of the SM contribution. However, this contribution
goes down very quickly when $\tan\beta$ increases. Given our approximation
($m_b\tan\beta\ll m_t\cot\beta$),
we cannot exceed $\tan\beta\sim 5$ and
still trust our results. Note that it has been shown in [24]
\footnote{$^5$}{the authors included the bottom quark
mass only for the diagram with two charged Higgs bosons
within the loop}, that this
contribution goes down very quickly when $\tan\beta$ is large; we certainly
see this trend.

In figs. 3 and 4, we show the chargino and gluino contributions. The global
behaviour is clear: for small gluino mass and small values of $m_S$, the
gluino contribution is important no matter what values the other parameters
have. On the other hand, for large gluino mass and/or large values of $m_S$,
the chargino contribution vastly dominates. The only exception to this rule is
for very large values of $\tan\beta$ ($\sim$ 30 or higher).
In this very special case,
the gluino contribution can be important, even for large gluino mass and
values of $m_S$ up to $400-500$\ GeV. This is due to the fact that such large
values of $\tan\beta$
can push down the mass of one of the scalar b-quark eigenstates; well below
the scalar top-quark eigenstates. Even requiring all scalar-quark masses to
be larger than the experimental limit of 92 GeV \footnote{$^6$}{this lower
limit is
model dependant, ref.[1]}
it is still possible for the gluino
contribution to be larger than the chargino contribution by a factor of
6 or so at the minium allowed value of $m_S$. We also note that values of
$\mu$ like 100 GeV or 400 GeV for a value of $m_{g_2}$ of 200 GeV will increase
(in magnitude) the contributions from the charginos while negative values
of $\mu$ will decrease them in magnitude. Furthermore, the effects of the
mixing of the scalar partners with the top and bottom quarks are more
important for
large values of $m_S$: the contributions from the charginos don't decrease as
quickly with the mixing. For small values of $m_S$, there is also an
enhancement.

The factor $c$ that enters the b-quark mixing matrix is also very important.
We find that reducing it from 1 to 1/2 reduces the gluino contributions by
a factor of $\sim 8$ for small values of $m_S~(\sim 200~GeV)$ while
it reduces them by a factor of $\sim 4$ for very large values of
$m_S~(\sim 1~TeV)$. The first factor will vary with $m_{g_2}$, $\mu$ and
$tan\beta$ but the
reduction of 4 for large $m_S$ is rather stable. Typically it will vary between
3.5 and 4.1. It can be understood by the fact that for those large scales, the
mass eigenvalues are almost insensitive to $c$ but the mixing angles are almost
linearly dependant on $c$.

Finally, one must not forget that in eq.(5) and eq.(6)
$K_{31}^\ast K_{33}$\ have not necessarily the same
values as in the SM.
This was shown in eq.(33) in [14]: the
Kobayashi--Maskawa matrix in the couplings of the
charginos to quarks and scalar quarks is multiplied by
another matrix $V_u$, which can be parametrized as follows:
$$V_u=\left(\matrix{1&\varepsilon_u&\varepsilon_u^2\cr
 -\varepsilon_u&1&\varepsilon_u\cr-
\varepsilon_u^2&-\varepsilon_u&1\cr}\right)
\eqno (8)$$
\noindent
so that $K\equiv V_u\cdot K_{SM}$. Clearly,
if $\epsilon\ll 1$ then $K\sim K_{SM}$.
However, with
$\varepsilon_u=0.5$\ $K^\ast_{31}K_{33}$\ is enhanced by a
factor of 14 over the SM value. This enhancement falls
quickly as $\varepsilon_u$\ decreases: to 3 for
$\varepsilon_u=0.3$\ and to $-0.2$\ for $\varepsilon_u=0.1$.
Note that because of the uncertainties in $K_{SM}$ this last
factor has a large error.
\hfill\break\indent
We have a similar matrix in the gluino--down quark--scalar
down quark couplings as was shown in eq.(29) in [14].
Here $K=V_d$\ where $V_d$\ is a similar matrix as $V_u$\
in eq.(8). For $\varepsilon_d=0.1$\ $K^\ast_{31}K_{33}$ is
identical to the SM values, whereas $\varepsilon_d=0.5$\
enhances it by 25 and $\varepsilon=0.3$\ by 9.
Considering that these values are to be squared in the
mass difference of the $B^0_D$\ system we can use that
enhancement to put limits on $\varepsilon_{u,d}$.
In the case at hand, $\varepsilon_u$\ has to be smaller than $0.2$\ and
$\varepsilon_d$\ smaller than $0.1$\ to keep the results
lower than the measured value of $\Delta m_{B^0_d}/m_{B^0_d}$.
This is not very constraining yet but it is already better than the limit one
can get from current data on rare Kaon decays [21].

\hfill\break\vskip.12cm\noindent
{\bf IV. CONCLUSIONS}\vskip.12cm
In this paper we presented the
contributions from charginos and scalar up quarks
as well as gluinos and scalar down quarks
to the mass difference in the $B_d^0$\
system via box diagrams. We gave exact results
and included the mixing of the charginos and
the mixing of the scalar top and bottom quarks. We
have shown that in the case of charginos and scalar
top quark the mixing becomes important and leads to
an enhancement of the results. Whereas in the case of
gluinos and scalar bottom quark its mixing is, as expected,
less important, even for higher values of $\tan\beta$\ the results
are dimisnished only by a few per cents.
We have shown that for reasonable values of the SUSY
parameters the contribution of the
box diagrams with charginos and scalar
up quarks can be of the same order as
those of the SM diagrams, but with opposite sign.
The same goes for the contribution of the gluino and
scalar down quarks box diagrams, which has  the same sign as
the SM contribution.
Since we have shown that despite the smallness of
the weak coupling constant compared to the strong coupling
constant charginos and scalar up quarks cannot be
neglected, we believe that for a complete analysis
it will be necessary also to include the diagrams with
neutralinos and scalar down quarks within the loops [26];
this is also supported by the fact that the lower mass bound of the smallest
neutralino by LEP data is only 30 GeV.
The contribution from the charged Higgs boson
to the mass
difference in the $B_d^0$\ system can be very important for small values
of $tan\beta$ and small Higgs masses. When the Higgs mass becomes large
($\sim$ 500 GeV) and/or $2\le tan\beta$, this contribution becomes small and
even negligible compared to the chargino contribution.
We presented
the results in the case where $m_b\tan\beta\ll m_t\cot\beta$.
An exact calculation including the bottom quark mass
therefore is highly desirable.
\hfill\break\indent
 A complete
analysis within the MSSM where all its particles
are taken within the relevant box diagram without
neglecting any mixing angles and mass eigenvalues
might therefore be of special interest and give
new bounds on SUSY parameters by the measured value
of $\Delta m_{B_d^0}/m_{B_d^0}$. Such a study becomes quite relevant in
the light of the upcoming {\it B-factories}.
\vfill\break\noindent
%\hfill\break\vskip.1cm\noindent
{\bf V. ACKNOWLEDGMENTS}\vskip.12cm
One of us (H.K.) would like to thank the physics department
of Carleton university
for the use of their computer
facilities.
The figures were done with the very user
friendly program PLOTDATA from TRIUMF.
\hfill\break\indent
This work was partially funded by funds from the N.S.E.R.C. of
Canada and les Fonds F.C.A.R. du Qu\'ebec.
\hfill\break\vskip.12cm\noindent
{\bf VI. APPENDIX A}\vskip.12cm
For the box diagram we have to calculate the following
integrals:
$$\eqalignno{F^{\mu\nu}_{abij}:=&\int{{d^4k}\over{(2\pi)^4}}
{{k^\mu k^\nu}\over{(k^2-m_a^2)(k^2-m_b^2)(k^2-m_i^2)(k^2-m_j^2)}}
&(A0)\cr
F^{\mu\nu}_{abij}=:&{{+ig^{\mu\nu}}\over{4(4\pi)^2}}
\tilde F^{ij}_{ab}\cr}$$
$m^2_i\not=m^2_j\not=m^2_a\not=m^2_b$
$$\eqalignno{\tilde F^{ij}_{ab}=-{1\over{(m_j^2-m_i^2)(m_b^2-
m_a^2)}}\bigl\lbrace&{1\over{(m_i^2-m_a^2)(m_i^2-m_b^2)}}
\lbrack m_i^4(m_b^2\ln{{m_b^2}\over{m_i^2}}-m_a^2\ln {{m_a^2}
\over{m_i^2}})\cr
&-m^2_im^2_am_b^2\ln{{m_b^2}\over{m_a^2}}\rbrack-
(m^2_i\leftrightarrow m^2_j)\bigr\rbrace&(A1)\cr}$$
$$\eqalignno{m_j^2\not=&m_i^2\not=m^2_a=m^2_b\cr
\tilde F^{ij}_{aa}=&-{1\over{(m_j^2-m_i^2)}}\bigl\lbrace
{1\over{(m_i^2-m_a^2)}}\lbrack m^2_a-{{m_i^4}\over{(m_i^2-
m_a^2)}}\ln{{m_i^2}\over{m_a^2}}\rbrack-(m_i^2\leftrightarrow
m_j^2)\bigr\rbrace&(A2)\cr
m_i^2=&m_j^2\not=m_a^2\not=m^2_b\cr
\tilde F^{ii}_{ab}=&\tilde F^{ij}_{aa}(m_a^2\leftrightarrow
m_i^2,\ m_b^2\leftrightarrow m_j^2)&(A3)\cr
m_j^2=&m_i^2\not= m_a^2=m_b^2\cr
\tilde F^{ii}_{aa}=&-{{(m_i^2+m_a^2)}\over{(m_i^2-m_a^2)^2}}
\lbrace 1-{{2m_i^2m_a^2}\over{(m_i^4-m_a^4)}}\ln{{
m_i^2}\over{m_a^2}}\rbrace &(A4)\cr}$$
The second integral is given by:
$$\eqalignno{_MF^{ij}_{ab}:=&\int{{d^4k}\over{(2\pi)^4}}
{{m_im_j}\over{(k^2-m_a^2)(k^2-m_b^2)(k^2-m_i^2)(k^2-m_j^2)}}
&(A5)\cr
_MF^{ij}_{ab}=:&{{-ig^{\mu\nu}}\over{(4\pi)^2}} _M\tilde F^{ij}
_{ab}\cr}$$
$m^2_i\not=m^2_j\not=m^2_a\not=m^2_b$
$$\eqalignno{_M\tilde F^{ij}_{ab}=-{{m_im_j}\over{(m_j^2-m_i^2)(m_b^2-
m_a^2)}}\bigl\lbrace&{1\over{(m_i^2-m_a^2)(m_i^2-m_b^2)}}
\lbrack m_i^2(m_b^2\ln{{m_b^2}\over{m_i^2}}-m_a^2\ln {{m_a^2}
\over{m_i^2}})\cr
&-m^2_am_b^2\ln{{m_b^2}\over{m_a^2}}\rbrack-
(m^2_i\leftrightarrow m^2_j)\bigr\rbrace&(A6)\cr}$$
$$\eqalignno{m_j^2\not=&m_i^2\not=m^2_a=m^2_b\cr
_M\tilde F^{ij}_{aa}=&-{{m_im_j}\over
{(m_j^2-m_i^2)}}\bigl\lbrace
{1\over{(m_i^2-m_a^2)}}\lbrack 1-{{m_i^2}\over{(m_i^2-
m_a^2)}}\ln{{m_i^2}\over{m_a^2}}\rbrack-(m_i^2\leftrightarrow
m_j^2)\bigr\rbrace&(A7)\cr
m_i^2=&m_j^2\not=m_a^2\not=m^2_b\cr
_M\tilde F^{ii}_{ab}=&_M\tilde F^{ij}_{aa}(m_a^2\leftrightarrow
m_i^2,\ m_b^2\leftrightarrow m_j^2,\ m_im_j
\rightarrow m_i^2)&(A8)\cr
m_j^2=&m_i^2\not= m_a^2=m_b^2\cr
_M\tilde F^{ii}_{aa}=&-{{m_i^2}\over{(m_i^2-m_a^2)^2}}
\lbrace 2+{{(m_i^2+m_a^2)}\over{(m_i^2-m_a^2)}}\ln{{
m_a^2}\over{m_i^2}}\rbrace&(A9)\cr}$$
\hfill\break\vskip.12cm\noindent
{\bf REFERENCES}\vskip.12cm
\item{[\ 1]} Review of Particle Properties, part 1,
Phys. Rev. {\bf D50}(1994)1172.
\item{[\ 2]} A.J. Buras, W. Slominski and H. Steger,
Nucl. Phys.{\bf B238}(1984)529, \hfill\break
 Nucl.Phys.{\bf B245}(1984)369
\item{[\ 3]}A.J. Buras, "Rare Deacys, CP Violation and
QCD", hep-ph/9503262.
\item{[\ 4]} V. Barger, T. Han and D.V. Nanopoulos,
 Phys. Lett.{\bf 194B}
(1987)312.
\item{[\ 5]} A. Ali, "$B^0-\overline B^0$\ mixing: A Reappraisal",
UCLA Workshop, 1987:110.
\item{[\ 6]}The CDF collaboration (F. Abe et al.),
Phys. Rev. Lett. {\bf 73}(1994)225.
\item{[\ 7]}H.E. Haber and G.L. Kane, Phys.Rep.{\bf 117}(1985)75.
\item{[\ 8]}J. Ellis and D.V. Nanopoulos, Phys.Lett.
{\bf 110B}(1982)44.
\item{[\ 9]}R. Barbieri and R. Gatto, Phys. Lett.
{\bf 110B}(1983)211.
\item{[10]}B.A. Campbell, Phys.Rev.{\bf D28}
(1983)209.
\item{[11]}H. K\"onig, Mod.Phys.Lett.{\bf A7}
(1992)279.
\item{[12]}H. K\"onig, Phys.Rev.{\bf D50}
(1994)3310.
\item{[13]}G. Couture, C. Hamzaoui and H. K\"onig,
"SUSY QCD correction to flavour changing top quark
decay", to be puplished in Phys.Rev.{\bf D}.
\item{[14]}M.J. Duncan, Nucl.Phys.
{\bf B221}(1983)285.
\item{[15]}J.F. Donoghue, H.P. Nilles and D. Wyler,
 Phys. Lett. {\bf 128B}
(1983)55.
\item{[16]}A. Bouquet, J. Kaplan and C.A. Savoy, Phys. Lett.
{\bf 148B}(1984)69.
\item{[17]}M.J. Duncan and J. Trampetic, Phys.Lett.{\bf 134B}
(1984)439.
\item{[18]} S. Bertolini, F. Borzumati and A. Masiero,
Phys. Lett{\bf 194B}(1987)551. Erratum: ibid{\bf 198B}(1987)590.
\item{[19]}G. Altarelli and P.J. Franzini,
Z.Phys.C--Part. and Fields {\bf 37}(1988)271.
\item{[20]}J.F. Gunion and H. Haber, Nucl.Phys.{\bf B272}
(1986)1.
\item{[21]}G. Couture and H. K\"onig,
"$K^+\rightarrow \pi^+\nu\overline\nu$\
and $K^0_L\rightarrow\mu^+\mu^-$\ decays
within the minimal supersymmetric standard model",
to be published in Z.Phys. {\bf C}.
\item{[22]}A. Djouadi, M. Drees and H. K\"onig,
Phys.Rev.{\bf D48}(1993)3081.
\item{[23]}T. Inami and C.S. Lim, Progr. Theor. Phys.{\bf 65}
(1981)297.
\item{[24]}G.C. Branco et al, Phys.Lett {\bf B337}(1994)316.
\item{[25]}Note that our result does not quite agree with that of Geng and
Ng, Phys. Rev. {\bf D38}, (1988) 2857; we would need a $-4$ instead of a $+4$
in
our equation to agree within an overall sign. This sign is numerically
important for small Higgs mass and $tan\beta\sim 1$; for
large Higgs mass ($\sim 300$ GeV or more) or
$2\le tan\beta$ the difference is not very large. There is also an overall 2/3
which comes from a different definition of $B_B$.
\item{[26]}G. Couture and H. K\"onig work in progress.
\hfill\break\vskip.12cm\noindent
%\vfil\vfil\eject
{\bf FIGURE CAPTIONS}\vskip.12cm
\item{Fig.1}The box diagrams with scalar up (down) quarks
and charginos (gluinos)
within the loop including the mass insertion diagram.
Note that in the case of the gluinos its arrow is the
other way round.
\item{Fig.2} The ratio of the total amplitude
$\Delta m_{B^0_d}^{\rm H^+}/\Delta m_{B^0_d}^{\rm SM}$\ as a function
 of $M_{H^+}$ for $tan\beta = 1$ (solid);
$tan\beta = 2$ (dash); $tan\beta = 5$ (dash-dot).
\item{Fig.3} The ratios
$\Delta m_{B^0_d}^{\rm Chargino}/\Delta m_{B^0_d}^{\rm SM}$\ and
$\Delta m_{B^0_d}^{\rm Gluino}/\Delta m_{B^0_d}^{\rm SM}$\
as a function of the scalar mass $m_S$ for $tan\beta = 1$ (solid);
$tan\beta = 2$ (dash); $tan\beta = 5$ (dash-dot); $tan\beta = 20$ (dot).
The negative values for large $m_S$ are the chargino contributions; those of
large amplitudes for small $m_S$ are the gluino contributions with
$m_{\tilde g} = 100~GeV$;
those of small amplitudes for small $m_S$ are the gluino contributions with
$m_{\tilde g} = 200~GeV$.
\item{Fig.4} The same as fig. 3 for different values of $m_{g_2}$ and $\mu$.
The curves of large amplitudes are the chargino contributions; those of smaller
amplitudes are the gluino contributions with $m_{\tilde g} = 100~GeV$; those of
smallest amplitudes are the gluino contributions with $m_{\tilde g} = 200~GeV$.

\vfill\break
\end
\bye